\documentclass[apj]{emulateapj} 

\usepackage{verbatim}

\newcommand{\msun}{$M_{\odot}$}

\newcommand{\pmpc}{Mpc$^{-1}$}
\newcommand{\kmps}{km~s$^{-1}$}

\newcommand{\oii}{[OII]$\lambda3727$~\AA}

\newcommand{\zwindowwide}{$0.6<z<0.9$}
\newcommand{\mcut}{10.25}
\newcommand{\ntot}{3326}   
\newcommand{\nmcut}{977} 
\newcommand{\ngf}{486} 
\newcommand{\mcutrange}{$M>1.8 \times 10^{10}$~\msun}
\newcommand{\mcutrangelog}{$\log M/M_{\odot} >~$\mcut}

\newcommand{\vjproj}{$C_{p}$}

\begin{document}

\title{The UVJ Selection of Quiescent and Star Forming Galaxies: Separating Early and Late-Type Galaxies and Isolating Edge-on Spirals\altaffilmark{1,2,3}}

\author{Shannon G. Patel\altaffilmark{4},
Bradford P. Holden\altaffilmark{5},
Daniel D. Kelson\altaffilmark{6},
Marijn Franx\altaffilmark{4},
Arjen van der Wel\altaffilmark{7},
Garth D. Illingworth\altaffilmark{5}
}

\altaffiltext{1}{This paper includes data gathered with the 6.5~meter Magellan Telescopes located at Las Campanas Observatory, Chile.}
\altaffiltext{2}{Based in part on data collected at Subaru Telescope and obtained from the SMOKA, which is operated by the Astronomy Data Center, National Astronomical Observatory of Japan.}
\altaffiltext{3}{Based on observations made with the NASA/ESA Hubble Space Telescope, obtained at the Space Telescope Science Institute. STScI is operated by the Association of Universities for Research in Astronomy, Inc. under NASA contract NAS 5-26555.}
\altaffiltext{4}{Leiden Observatory, Leiden University, P.O. Box 9513, NL-2300 AA Leiden, Netherlands; patel@strw.leidenuniv.nl}
\altaffiltext{5}{UCO/Lick Observatory, University of California, Santa Cruz, CA 95064}
\altaffiltext{6}{Observatories of the Carnegie Institution of Washington, Pasadena, CA 91101}
\altaffiltext{7}{Max-Planck-Institut f\"{u}r Astronomie, K\"{o}nigstuhl 17, D-69117, Heidelberg, Germany}

\begin{abstract}
We utilize for the first time HST ACS imaging to examine the structural properties of galaxies in the rest-frame $U-V$ versus $V-J$ diagram (i.e., the $UVJ$ diagram) using a sample at \zwindowwide\ that reaches a low stellar mass limit (\mcutrangelog).  The use of the $UVJ$ diagram as a tool to distinguish quiescent galaxies from star forming galaxies (SFGs) is becoming more common due to its ability to separate red quiescent galaxies from reddened SFGs.  Quiescent galaxies occupy a small and distinct region of $UVJ$ color space and we find most of them to have concentrated profiles with high S\'{e}rsic indices ($n>2.5$) and smooth structure characteristic of early-type systems.  SFGs populate a broad, but well-defined sequence of $UVJ$ colors and are comprised of objects with a mix of S\'{e}rsic indices.  Interestingly, most $UVJ$-selected SFGs with high S\'{e}rsic indices also display structure due to dust and star formation typical of the $n<2.5$ SFGs and late-type systems.  Finally, we find that the position of a SFG on the sequence of $UVJ$ colors is determined to a large degree by the mass of the galaxy and its inclination.  Systems that are closer to edge-on generally display redder colors and lower \oii\ luminosity per unit mass as a consequence of the reddening due to dust within the disks.  We conclude that the two main features seen in $UVJ$ color space correspond closely to the traditional morphological classes of early and late-type galaxies.
\end{abstract}


\section{Introduction}

\begin{figure*}
\epsscale{1}
\plotone{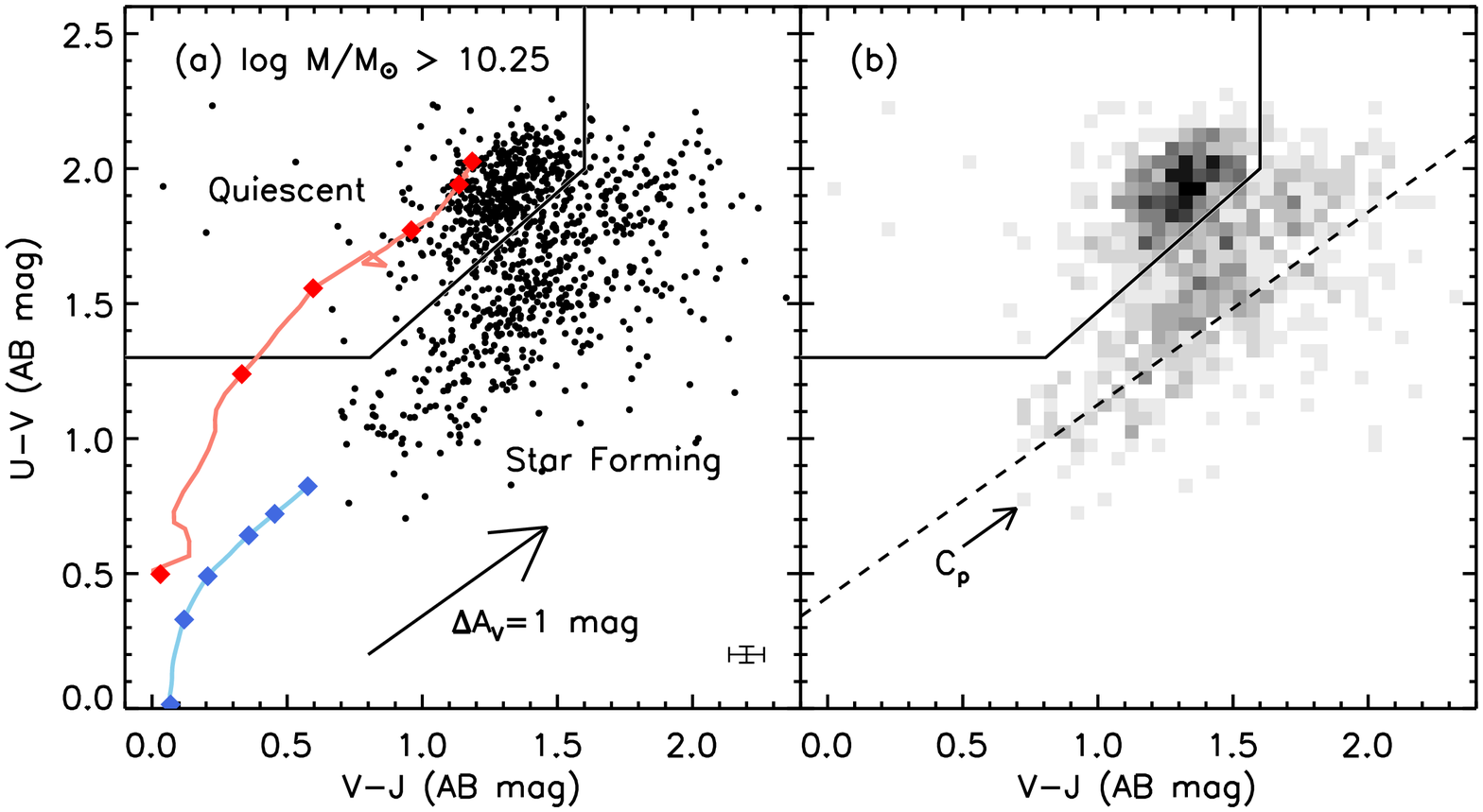}
\caption{($a$) Rest-frame $U-V$ vs. $V-J$ colors for \nmcut\ galaxies with mass \mcutrangelog\ at \zwindowwide.  The typical $\pm 1$-$\sigma$ statistical uncertainty in the rest-frame colors is indicated in the bottom right.  The \citet{williams2009} boundary (black wedge) separates quiescent galaxies (top left) from SFGs (bottom right).  The red line indicates the evolution of a BC03 solar metallicity single burst, while the blue line represents a constant SFR model (CSF).  The diamonds on each model track represent time steps of 0.1, 0.5, 1, 2, 3, and 5 Gyr (from bottom to top).  The arrow indicates the reddening vector.  ($b$) Gray scale representation of the density of points in panel ($a$).  The concentration of quiescent galaxies in the $UVJ$ plane is distinct from the broader distribution of SFGs.  The dashed line runs through the $t=5$~Gyr CSF time step and is parallel to the reddening vector.  We examine axis ratios and \oii\ luminosities along this line, \vjproj, in Figure~\ref{fig_boa_VJproj} as we hypothesize that reddening is a major factor in explaining the sequence of $UVJ$ colors for SFGs.} \label{fig_UVJ_letter}
\end{figure*}

In the absence of morphological information to distinguish early and late-type galaxies, the use of UV-optical colors has emerged as a commonly employed proxy in studies of galaxy evolution.  This practice is bolstered by the color bimodality \citep[e.g.,][]{strateva2001,baldry2004}.  For example, galaxies are often divided in a color-magnitude or color-mass diagram into red and blue, and the evolution in the luminosity or mass function of each color type tracked over cosmic time \citep[e.g.,][and several others]{bell2004b,faber2007,brown2007}.  The underlying assumption in the method above is that the red galaxies represent quiescent systems with little or no ongoing star-formation (SF).  However, it is well known that a reddened star forming galaxy (SFG) can display colors coincident with those on the red sequence \citep[e.g.,][]{maller2009}.  Selecting red galaxies based on a single rest-frame color, in combination with a magnitude or stellar mass, therefore results in a selection of both quiescent galaxies {\em and} reddened SFGs.  This presents a predicament for studies that use such a selection in order to quantify the evolution in the number of passive systems.

In this Letter, we highlight the utility of a color-color diagram, specifically rest-frame $U-V$ vs. $V-J$ (hereafter referred to as the $UVJ$ diagram), in distinguishing quiescent galaxies from SFGs, including those SFGs that are heavily reddened.  The $UVJ$ diagram is gaining increasing visibility in studies of galaxy evolution \citep[e.g.,][]{labbe2005,wuyts2007,williams2009,brammer2009,williams2010,whitaker2010,patel2011,quadri2011,brammer2011}.  In \citet{patel2011} we hinted at the possibility of the $UVJ$ selection of quiescent and SFGs being used to also distinguish morphologically early and late-type systems.  Here we introduce for the first time an analysis based on {\em HST} ACS imaging and examine the structural properties of galaxies in $UVJ$ color space in order to address the prospects of using two rest-frame colors in classifying both the recent SFHs and morphologies of galaxies.

We use a stellar mass-limited sample of galaxies at \zwindowwide\ to carry out our analysis.  As is well known, many galaxy properties correlate with stellar mass and it is therefore crucial that one control for stellar mass in interpreting results.  The redshift range examined here provides a suitable case study for a sample of galaxies in the distant universe.  The rest-frame $J$-band magnitudes at these redshifts represent observed $K_s$, which is typically the reddest bandpass accessible from the ground for deep wide-field galaxy surveys.

We assume a cosmology with $H_0=70$~\kmps~\pmpc, $\Omega_M = 0.30$, and $\Omega_{\Lambda} = 0.70$.  Stellar masses are based on a Chabrier IMF \citep{chabrier2003}.  All magnitudes are reported in the AB system.

\section{Data and Analysis}

\begin{figure*}[t]
\epsscale{0.9}
\plotone{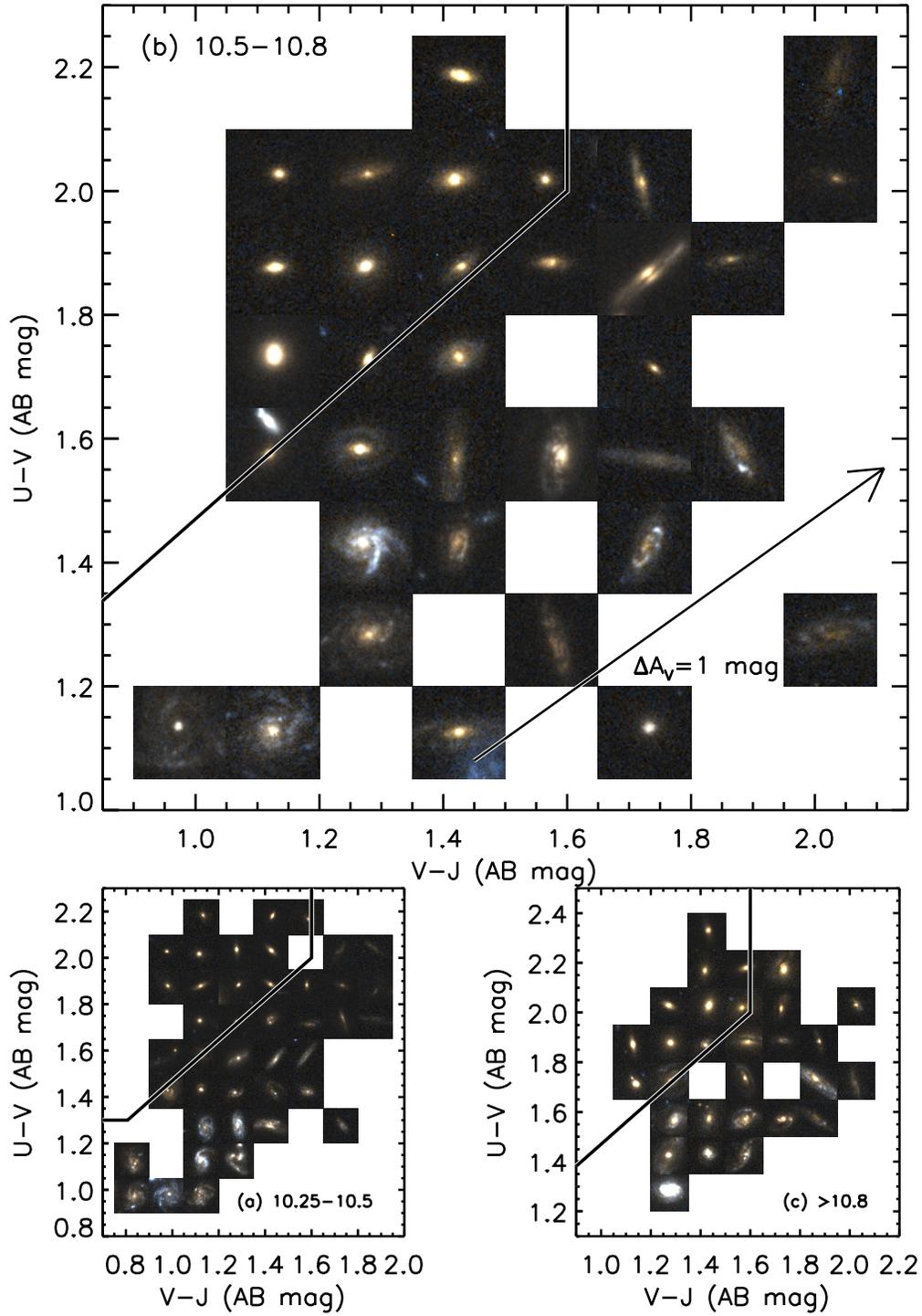}
\caption{Rest-frame $UVJ$ diagram at \zwindowwide\ represented with {\em HST} ACS color postage stamps (either F606W and F814W or F625W and F775W, 3\arcsec\ on a side) for galaxies in three stellar mass bins: ($a$) $10.25<\log(M/M_{\odot})<10.5$, ($b$) $10.5<\log(M/M_{\odot})<10.8$, ($c$) $\log(M/M_{\odot})>10.8$.  In all mass bins, $UVJ$-selected quiescent galaxies are mostly early-type systems while SFGs are primarily late-type systems.  Bluer SFGs are generally face-on disks, while redder ones tend to be viewed edge-on.} \label{fig_UVJ_ACS_binmass}
\end{figure*}

Data products for our sample are discussed in detail in \citet{patel2009,patel2009b,patel2011} and briefly summarized here.  Ground based imaging consists of Magellan IMACS $B$, Subaru Suprime-Cam $VRiz$, and Du Pont WIRC $K_s$.  The $z$-band was used for spectroscopic selection ($18<z_{\rm AB}<23.3$~mag).  We obtained spectroscopy with IMACS on Magellan, using a low-dispersion prism (LDP) in place of the grating.  Roughly $\sim 10,000$ redshifts were measured over a $\sim 0.2$~deg$^2$ field.  The overall spectroscopic completeness is $\sim 74\%$.  Our sample in this work is restricted to galaxies in the redshift range \zwindowwide, with the lower redshift cutoff determined by the bright magnitude limit of our survey and the upper redshift cutoff determined by the desired mass cut and with some consideration for the redshift precision.  In this redshift interval, the LDP redshifts have a precision of $\sigma_z/(1+z) \approx 0.012$ when compared to redshifts obtained from higher resolution spectroscopy.  We find \ntot\ galaxies in this redshift interval with $z_{\rm AB}<23.3$~mag.  We fit \citet[][hereafter referred to as BC03]{bc03} $\tau$-models to SEDs comprised of both the broadband photometry and LDP spectra in order to extract relevant rest-frame quantities such as stellar masses.  The limiting stellar mass for constructing representative samples of both star-forming and passive galaxies is determined by the latter and is \mcutrange\ (\mcutrangelog) at $z=0.9$.  The SED fitting procedure also provides measurements of \oii\ luminosities.  Rest-frame magnitudes and colors were computed with the ground-based imaging and using the $K$-correction technique described in \citet{rudnick2003}.

We use {\em HST} ACS imaging \citep{ford2003} to measure the structural properties of galaxies.  About $\sim 50\%$ of our mass-limited sample with $K_s$ imaging has ACS coverage in either the F775W or F814W filter.  The color postage stamps shown in the next section are for the subset of our sample with ACS coverage in two filters (i.e., either F606W and F814W, or F625W and F775W).  We used GALFIT \citep{peng2002} on the F775W or F814W imaging to measure axis ratios, $b/a$, S\'{e}rsic indices, $n$, and effective radii, $R_e$.  We direct the reader to \citet{holden2009} for a detailed discussion on the fitting procedure as well as results from simulations pertaining to the reliability of the axis ratio measurements.  We also measured the bumpiness parameter, $B$, which provides an estimate of the higher order structure within galaxies and is defined as the ratio of the rms of the residual from the GALFIT model fit and the model mean \citep[see, e.g,][]{blakeslee2006,vanderwel2008}.

Our final mass-limited sample at \zwindowwide\ with $K_s$ imaging and measured structural parameters from {\em HST} imaging is comprised of \ngf\ galaxies.

\bigskip
\bigskip
\bigskip

\begin{figure*}
\epsscale{1.2}
\plotone{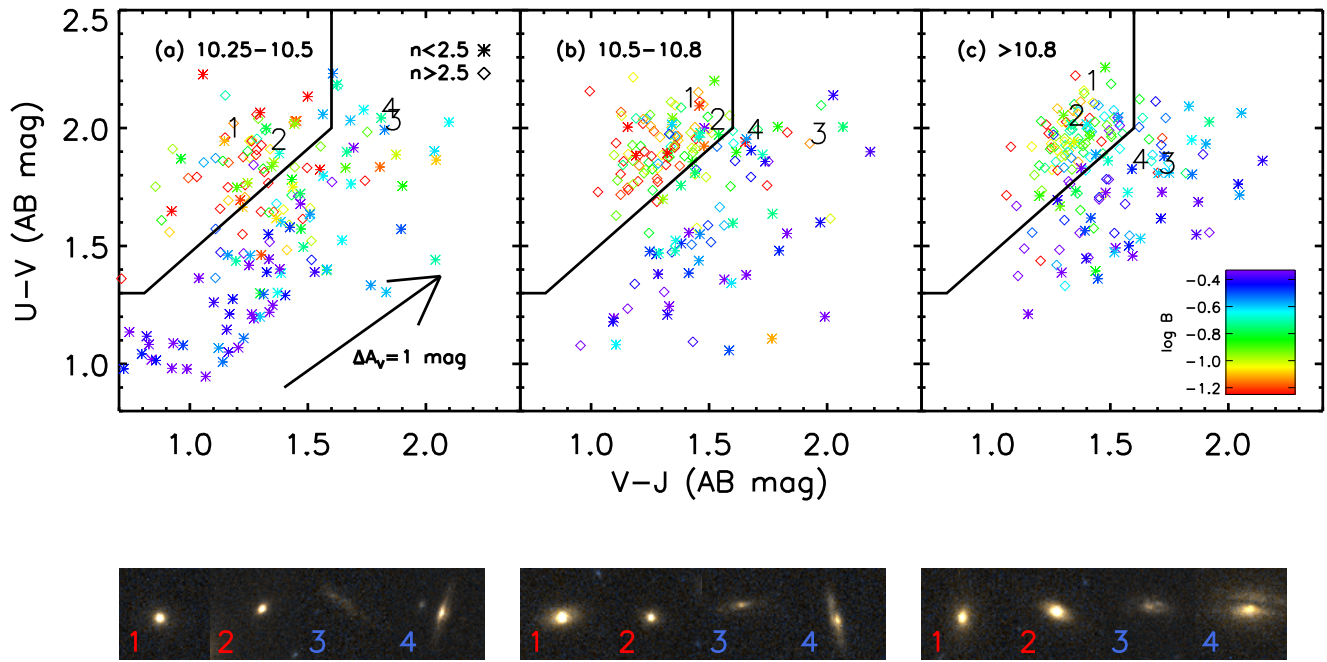}
\caption{Rest-frame $UVJ$ diagram for galaxies in three mass bins at \zwindowwide\ with structural properties measured from {\em HST} ACS imaging.  Asterisks indicate objects with S\'{e}rsic index $n<2.5$ while diamonds indicate $n>2.5$.  Data points are color-coded according to the bumpiness parameter, $B$, which measures the amount of higher order structure based on rms of the GALFIT residual.  The quiescent region of the $UVJ$ diagram is dominated by objects with high S\'{e}rsic indices and smooth profiles based on their low $B$ values.  The star-forming region is comprised of a mix of objects with low and high S\'{e}rsic indices.  However, SFGs with high S\'{e}rsic indices generally have high $B$ values suggestive of dust and/or SF, typical of late-type systems.  Below each $UVJ$ diagram are example color postage stamps for two quiescent and two SFGs with red optical colors ($U-V>1.8$~AB mag).  Each galaxy is labeled in the top panels.  The $UVJ$ selection is seen here to be effective in distinguishing early and late-type galaxies, even on the red sequence.} \label{fig_UVJ_sersic_bumpiness}
\end{figure*}

\section{The $UVJ$ Diagram}

\subsection{Selecting Quiescent and SFGs}

In Figure~\ref{fig_UVJ_letter} we present the rest-frame $UVJ$-diagram for a stellar mass-limited sample at \zwindowwide.  The combination of these two colors allow one to break the degeneracy between age and reddening for red galaxies as discussed below.  Quiescent galaxies are delineated from SFGs using the boundary computed by \citet{williams2009}.  We note that this boundary results in $\sim 31\%$ of galaxies with $U-V>1.7$~AB mag being classified as SFGs.  This fraction does not vary significantly for different mass bins above our mass cut.  Figure~\ref{fig_UVJ_letter}$b$ shows a gray scale representation of the density of points in the $UVJ$ diagram.  The concentration of quiescent galaxies is distinct from the broader distribution of $UVJ$ classified SFGs.

Model star formation histories (SFHs) reinforce the \citet{williams2009} division between quiescent and SFGs in the $UVJ$ diagram.  The color evolution of a BC03 solar metallicity single burst (SSP) and constant star formation rate model (CSF) are shown in Figure~\ref{fig_UVJ_letter}$a$.  After $\sim 3-5$~Gyr, the SSP lies in the region of the $UVJ$ diagram populated by quiescent galaxies.  Meanwhile, after $\sim 5$~Gyr, the CSF lies in the star forming region of the $UVJ$ diagram where no galaxies with mass \mcutrange\ can be found.  However, the reddening vector, computed assuming a \citet{calzetti2000} reddening law, suggests that the addition of varying amounts of extinction can explain a large range of colors for SFGs above our mass limit.  \citet{patel2011} show the color evolution of more complex SFHs in the $UVJ$ diagram, including bursts and other composite SFHs.  We reiterate their finding that the slightest amount of star formation (SF) on top of a dominant older stellar population can easily move galaxies from the quiescent region to the star forming region.  The SSP and CSF models shown here are intended to provide a simple overview of the kinds of SFHs that can produce colors consistent with $UVJ$ classified quiescent and SFGs.

Finally, \citet{patel2011} showed that MIPS 24~\micron\ detections primarily trace the region of the $UVJ$ diagram occupied by SFGs including those with the reddest colors ($U-V>1.7$~AB mag), therefore providing empirical support for the \citet{williams2009} boundary.

\subsection{The Morphological Structure of Galaxies in the $UVJ$ Diagram}

With {\em HST} ACS imaging, we gain a new understanding for the kinds of galaxies that populate various regions of the $UVJ$ diagram.  These data provide perhaps the most informative explanation as to why the $UVJ$ selection works as well as it does.  In Figure~\ref{fig_UVJ_ACS_binmass}, for three different mass bins, we show the $UVJ$ diagram populated with {\em HST} ACS color postage stamps for the subset of galaxies with ACS coverage.  Each color-color bin (0.15~mag in size) is represented with a postage stamp of a random galaxy within the bin.  These images display rest-frame light at $\sim 3000-5000$~\AA, similar to the wavelength range probed by the $U-V$ color.  The contrast in structure between galaxies selected in the $UVJ$ diagram to be quiescent versus star forming is striking.  The quiescent region is populated primarily by early-type galaxies.  In contrast, the star forming region is dominated by late-type galaxies.  

In Figure~\ref{fig_UVJ_sersic_bumpiness}, we look more quantitatively at the structural properties of galaxies in the $UVJ$ diagram for the same three mass bins as Figure~\ref{fig_UVJ_ACS_binmass}.  The symbols indicate whether galaxies have $n<2.5$ (asterisks) or $n>2.5$ (diamonds) and are color-coded according to the bumpiness value, $B$.  Galaxies with higher order structure, suggestive of dust and/or SF in late-type systems, have higher values of $B$.  

In the quiescent region of the $UVJ$ diagram, the fraction of galaxies with $n>2.5$ for the low, intermediate, and high mass bin is $68 \pm 6\%$, $85 \pm 4\%$, and $89 \pm 3\%$, respectively.  We note that most of the quiescent galaxies with $n<2.5$ generally have low $B$ values, indicating featureless profiles.  The somewhat higher fraction at lower masses generally reside in more over-dense environments relative to $n<2.5$ SFGs, and could therefore represent quenched satellites.  Thus, in all mass bins, the quiescent region is indeed dominated by galaxies with concentrated and/or featureless surface brightness profiles typical of early-type galaxies.  

In the star-forming region of the $UVJ$ diagram, the proportion of galaxies that have $n<2.5$ for the low, intermediate, and high mass bin is $75 \pm 4\%$, $58 \pm 6\%$, and $48 \pm 6\%$, respectively.  Thus, a substantial fraction of SFGs are not best-fit with low S\'{e}rsic indices indicative of disk-dominated systems, especially at higher masses.  However, for those SFGs with $n>2.5$, most also have high values for $B$ indicating the presence of higher order structure common to late-type galaxies.  These galaxies generally have strong bulges, likely responsible for driving the S\'{e}rsic fit, but also a visually identifiable disk component with strong dust and/or SF, a key difference from the $n>2.5$ quiescent systems.  Thus, the quiescent and star forming regions of the $UVJ$ diagram have been shown here to correspond closely to the traditional morphological classes of early and late-type galaxies.

\section{Inclination: A Key Source for the Sequence of $UVJ$ Colors of SFG\lowercase{s}} \label{sec_inclination}

\begin{figure}
\epsscale{1.2}
\plotone{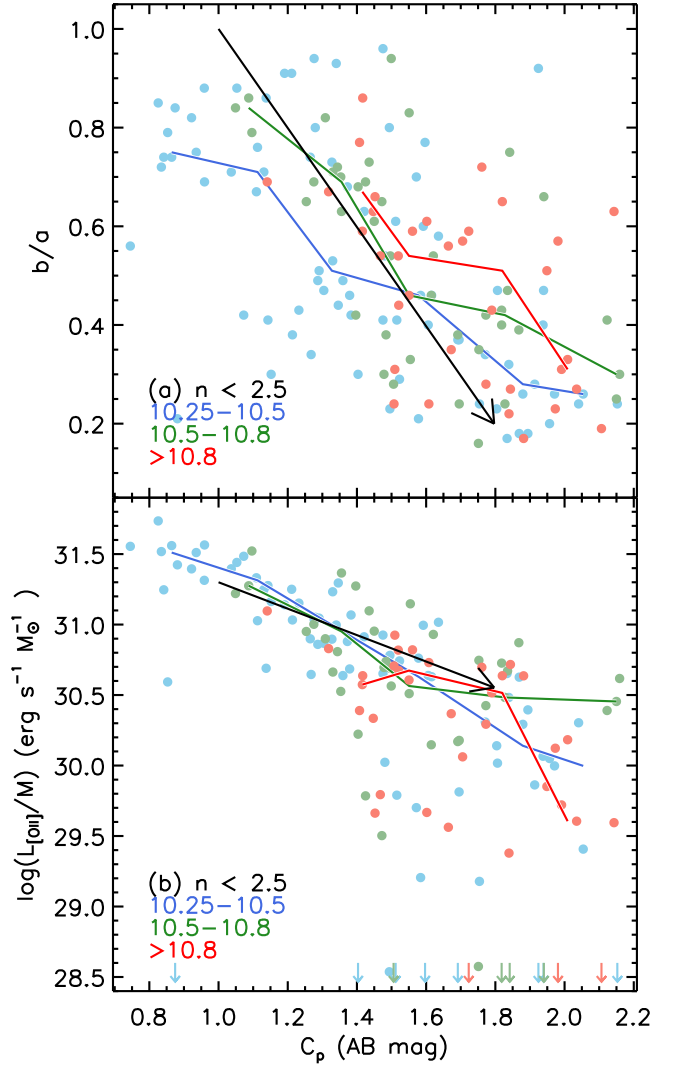}
\caption{($a$) Axis-ratio, $b/a$, versus \vjproj\ for $UVJ$-selected SFGs at \zwindowwide\ with S\'{e}rsic index $n<2.5$.  The quantity \vjproj\ is the $V-J$ color of a galaxy after projecting it onto the reddening vector seen in Figure~\ref{fig_UVJ_letter}$b$ (dashed line in that figure).  Objects are color-coded according to the same mass bins as in Figure~\ref{fig_UVJ_ACS_binmass}.  The solid colored lines indicates the median value of $b/a$ as a function of \vjproj.  A clear correlation exists between $b/a$ and \vjproj.  SFGs with bluer \vjproj\ colors have axis-ratios closer to unity indicating that they are viewed close to face-on, while those with red colors have small axis ratios indicating that they are viewed closer to edge-on.  ($b$) \oii\ luminosity, normalized by stellar mass, versus \vjproj\ for the same sample as in panel ($a$).  Arrows on the bottom indicate objects with a lack of \oii\ emission.  SFGs with redder \vjproj\ colors have less \oii\ emission, consistent with a scenario in which the light is extincted by dust in disks due to the closer to edge-on orientation.  The long black arrows in panels ($a$) and ($b$) indicate the impact on \vjproj\ and $L_{\rm [OII]}/M$, respectively, when going from face-on to edge-on, based on simulations by \citet{rocha2008}.} \label{fig_boa_VJproj}
\end{figure}

As previously noted, SFGs span a wide range of colors in the $UVJ$ diagram.  There are two possibilities for the source of this diversity in the $UVJ$ colors at a fixed mass: (1)  differences in the amount of reddening, and/or (2) differences in the SFHs.  Metallicity is not expected to be a significant factor, given that at a fixed mass, the variation in metallicity is only $\sim \pm 0.1$~dex \citep{tremonti2004}.  For a CSF model, this variation in metallicity would result in a negligible change in the $U-V$ color and a change of only $\sim \pm 0.1$~mag in the $V-J$ color, which is not enough to explain the broad range of $UVJ$ colors for SFGs.  We focus on option (1) above in this Letter and will show that this component can account for a significant portion of the variation in $UVJ$ colors for SFGs.  In fact, Figure~\ref{fig_UVJ_ACS_binmass} hints at a trend between the viewing inclination of SFGs, which is related to the degree of reddening, and $UVJ$ colors.  While bluer SFGs appear to be viewed closer to face-on, redder SFGs are more likely to be viewed closer to edge-on.  This is apparent for all mass bins.

Figure~\ref{fig_boa_VJproj} reinforces the interpretation that the colors of SFGs are driven to a large degree by the viewing angle.  In this figure, we limit our analysis to $n<2.5$ SFGs because the impact due to inclination is easier to gauge on these systems that are more likely to be disk-dominated.  Figure~\ref{fig_boa_VJproj}$a$ shows the axis ratio, $b/a$, vs. \vjproj, for SFGs with $n<2.5$.  The axis ratio serves as a proxy for the inclination.  The quantity \vjproj\ is simply the value of $V-J$ a SFG would have after projecting its $U-V$ and $V-J$ colors onto a reddening vector that runs through the $t=5$~Gyr CSF time step (dashed black line in Figure~\ref{fig_UVJ_letter}$b$).  In simpler terms, we are examining how the properties of SFGs vary along the reddening vector.  For $n<2.5$ systems, Figure~\ref{fig_boa_VJproj}$a$ shows a clear correlation between $b/a$ and \vjproj\ for galaxies in all mass bins.  The Pearson correlation coefficients for the low, intermediate, and high mass bin are $-0.58 \pm 0.08$, $-0.63 \pm 0.09$, and $-0.50 \pm 0.13$ respectively, where the uncertainties are computed by bootstrapping.  These values suggest a strong correlation between the two quantities.

We use the models by \citet{rocha2008} to estimate the reddening in $U-V$ and $V-J$ resulting from inclining a dusty disk.  The results for their Sbc model galaxy at $b/a=0.2$ (inclination, $i \approx 80^{\circ}$) implies $E(U-V) \approx 0.50$~mag and $E(V-J) \approx 0.62$~mag relative to a face-on orientation or equivalently $\sim 0.80$~mag for \vjproj.  This is comparable to the observed spread in \vjproj\ seen in Figure~\ref{fig_boa_VJproj}$a$ (black arrow) but slightly higher than the values implied for low redshift SFGs studied by \citet{yip2010b}.  The reddening due to dust could simply be lower in disks at low redshift.

Since SFGs with redder \vjproj\ are more likely to be viewed closer to edge-on, one would expect light from star forming regions to be extincted within such highly inclined disks.  Figure~\ref{fig_boa_VJproj}$b$ shows \oii\ luminosities, normalized by stellar mass, versus \vjproj\ for the same sample of SFGs as in Figure~\ref{fig_boa_VJproj}$a$.  The \oii\ luminosities decline at redder \vjproj\ at a fixed mass by roughly an order of magnitude.  This finding provides further support for the interpretation that the $UVJ$ colors of disk-dominated SFGs are driven by the viewing angle.  If the decline in \oii\ luminosity with \vjproj\ for SFGs was instead driven by a drop in the SFR, we would not expect to find large numbers of such galaxies in the star forming region of the $UVJ$ diagram.  For example, a factor of $\sim 6$ drop in the SFR of a CSF model at $t=5$~Gyr, would result in such a galaxy moving into the quiescent region after only $\sim 300$~Myr.  Additionally, the predicted decline in $L_{\rm [OII]}/M$ due to inclination from the \citet{rocha2008} models follows the data quite well (black arrow).

The correlations between $b/a$ or $L_{\rm [OII]}/M$ and \vjproj\ is almost as strong at the highest masses, suggesting that inclination is an important factor in that mass regime as well.  However, it is worth noting that the highest mass bin lacks objects that populate the bluer star forming regions of the $UVJ$ diagram (see Figure~\ref{fig_UVJ_sersic_bumpiness}), where lower mass objects can be found.  This suggests less vigorous SF in higher mass SFGs (i.e., lower specific SFR), in accord with the findings of other works \citep[e.g.,][]{brinchmann2004,noeske2007b}, and also possibly higher levels of dust and therefore reddening in these galaxies relative to lower mass ones  \citep[see, e.g.,][]{garn2010b} when viewed face-on.

The spread in $UVJ$ colors for $n>2.5$ SFGs is also significant.  However, their colors do not show a strong correlation with axis ratio.  Visual inspection of these systems suggests strong dust in the redder $n>2.5$ SFGs.  Thus, the dust geometry is possibly more complex or the inclination is not easily connected to the axis ratio (as can be the case for triaxial systems), leading to a poor correlation between $b/a$ and \vjproj.  We note however, that \citet{driver2007b} find strong inclination dependent attenuation in nearby bulges.

\section{Summary}

In the rest-frame $UVJ$ diagram, quiescent galaxies occupy a concentrated region of color space that is distinct from the sequence of $UVJ$ colors of SFGs, which extends onto the red sequence.  Utilizing {\em HST} ACS imaging for the first time to study the structural parameters of galaxies in the $UVJ$ diagram at \zwindowwide, we find that the commonly employed $UVJ$ selection of quiescent and SFGs corresponds closely to the distinction between traditional morphological classes of early and late-type galaxies.  Meanwhile, the position of a SFG along the extended sequence of $UVJ$ colors is determined to a large degree by its stellar mass and viewing inclination.  Systems that are closer to edge-on generally display redder $UVJ$ colors and have lower \oii\ luminosity per unit mass as a consequence of the reddening from dust within the inclined disks.  Thus, most of the $UVJ$-selected SFGs on the optical red sequence are highly inclined spirals.  The use of only two rest-frame colors in classifying both the recent SFHs and morphologies of galaxies will be invaluable for future studies of galaxy evolution.

\acknowledgements
We wish to acknowledge those who have contributed to the construction and deployment of IMACS as well as Scott Burles for developing the low-dispersion prism, and the PRIMUS collaboration for allowing us to investigate galaxies with their hardware.  This research was supported by an NWO-Spinoza Grant.




\end{document}